\documentclass[12pt]{article}
\usepackage{amsfonts,bm}
\usepackage{graphicx}
\baselineskip
\offinterlineskip
\begin{document}

\begin{center}
{\bf Hamilton Operators, Discrete Symmetries, Brute Force and
SymbolicC++}
\end{center}

\begin{center}
{\bf  Willi-Hans Steeb$^\dag$ and Yorick Hardy$^\ast$} \\[2ex]

$\dag$
International School for Scientific Computing, \\
University of Johannesburg, Auckland Park 2006, South Africa, \\
e-mail: {\tt steebwilli@gmail.com}\\[2ex]

$\ast$
Department of Mathematical Sciences, \\
University of South Africa, Pretoria, South Africa, \\
e-mail: {\tt hardyy@unisa.ac.za}\\[2ex]
\end{center}

\strut\hfill

{\bf Abstract} To find the discrete symmetries of
a Hamilton operator $\hat H$ is of central importance in
quantum theory. Here we describe and implement
a brute force method to determine the discrete symmetries 
given by permutation matrices for Hamilton operators
acting in a finite-dimensional Hilbert space.
Spin and Fermi systems are considered as examples.
A computer algebra implementation in SymbolicC++ is provided.

\strut\hfill

\section{Introduction}

In quantum mechanics the system is described by a 
self-adjoint (Hamilton) operator $\hat H$ acting in
a Hilbert space $\cal H$. Here we consider the finite
dimensional Hilbert space ${\mathbb C}^n$ where $\hat H$
is a hermitian matrix \cite{1,2,3,4,5}.
One of the main tasks is to find the $n \times n$ unitary 
matrices $U$ such that $U^*\hat HU=\hat H$, where $U^*=U^{-1}$.
The $n \times n$ unitary matrices form the compact Lie
group $U(n)$. Note that if $U^*\hat HU=\hat H$ and $V^*\hat HV=\hat H$
then $(UV)^*\hat H(UV)=\hat H$, where $(UV)^* \equiv V^*U^*$.
Thus the set of matrices that keep $\hat H$ invariant form
a group themselves \cite{6}. 
\newline

An important finite subgroup of the group $U(n)$ are the $n \times n$ permutation 
matrices. The number of $n \times n$ permutation matrices is $n!$. 
For a given hermitian $n \times n$ matrix $\hat H$ we want to find all
the $n \times n$ permutation matrices $P$ such that
$$
P^T \hat H P=\hat H
$$ 
where $P^{-1}=P^T$. These permutation matrices form a subgroup of all 
$n \times n$ permutation matrices. Obviously the $n \times n$ identity 
matrix $I_n$ satisfies $I_n\hat HI_n=\hat H$. 
Here we describe and implement in SymbolicC++ a brute force
method to find these permutation matrices.
\newline

After this finite group has been found one determines the conjugacy classes. Now for a finite 
group $G$ the number of conjugacy classes is equivalent to the number of
non-equivalent irreducible matrix representations. From the conjugacy 
classes and the permutation matrices we can construct projection
matrices to decompose the Hilbert space into invariant sub Hilbert
spaces \cite{6}. For example, if the permutation matrix $P$ satisfies $P^2=I_n$, then
$\Pi_1=(I_n+P)/2$, $\Pi_2=(I_n-P)/2$
are projection matrices (with $\Pi_1\Pi_2=0$) which can be
utilized to decompose the Hilbert space ${\mathbb C}^n$
into invariant subspaces.
 
\section{Examples} 

We consider four examples: two Fermi systems and two spin systems.
\newline

Example 1. Let $c_j^\dagger$, $c_j$ $(j=1,2,3)$ be Fermi creation
and annihilation operators, respectively. Consider the Hamilton 
operator
$$
\hat H = t(c_1^\dagger c_2 + c_2^\dagger c_1 + c_2^\dagger c_3 +
c_3^\dagger c_2 + c_1^\dagger c_3 + c_3^\dagger c_1) +
k_1 c_1^\dagger c_1 + k_2 c_2^\dagger c_2 + k_3 c_3^\dagger c_3 
$$
and the number operator $\hat N=c_1^\dagger c_1+c_2^\dagger c_2+c_3^\dagger c_3$.
Then $[\hat H,\hat N]=0$. Since $[\hat H,\hat N]=0$
we find $\hat N$ is a constant of motion, i.e. the total number of Fermi
particles remains constant in the sense that if $|n\rangle$ is an
eigenstate of the number operator $\hat N$ with eigenvalue $n$ at time
0, then $|n\rangle(t)=e^{-i\hat Ht/\hbar}|n\rangle$
remains an eigenstate of $\hat N$ with eigenvalue $n$ for all times.
Given a basis with two Fermi particles
$$
c_1^\dagger c_2^\dagger|{\bf 0}\rangle, \quad 
c_1^\dagger c_3^\dagger|{\bf 0}\rangle, \quad 
c_2^\dagger c_3^\dagger|{\bf 0}\rangle\,.
$$
Then we find the matrix representation of $\hat H$
$$
\hat H =
\pmatrix { k_1+k_2 & t & -t \cr t & k_1+k_3 & t \cr -t & t & k_2+k_3 }\,.
$$
The matrix representation of $\hat N$ is the diagonal matrix
$2I_3$, where $I_3$ is the $3 \times 3$ identity matrix. For 
$k_1 \ne k_2$, $k_1 \ne k_3$, $k_2 \ne k_3$ no non-trivial symmetry
is found. Also for $k_1 \ne k_2$, $k_1 \ne k_3$, $k_2=k_3$ no
non-trivial symmetry is found. For $k=k_1=k_2=k_3$ we obtain the permutation
matrix
$$
P = \pmatrix { 0 & 0 & 1 \cr 0 & 1 & 0 \cr 1 & 0 & 0 }\,.
$$
Thus using the projection matrices $\Pi_1=(I_3-P)/2$, 
$\Pi_2=(I_3-P)/2$ the Hilbert space ${\mathbb C}^3$ can be
decomposed into invariant subspaces. For the case $k=k_1=k_2=k_3$
we find the eigenvalues $2k-2t$ and $2k+t$ (twice).
\newline

Example 2. Consider the Hamilton operator (two-point Hubbard model)
$$
\hat H = t(c_{1\uparrow}^\dagger c_{2\uparrow} + 
c_{1\downarrow}^\dagger c_{2\downarrow} +
c_{2\uparrow}^\dagger c_{1\uparrow} + 
c_{2\downarrow}^\dagger c_{1\downarrow}) +
U(n_{1\uparrow} n_{1\downarrow} + n_{2\uparrow} n_{2\downarrow})
$$
where $n_{j\uparrow}:=c_{j\uparrow}^\dagger c_{j\uparrow}$, 
$n_{j \downarrow}:=c_{j\downarrow}^\dagger c_{j\downarrow}$. 
The operators $c_{j\uparrow}^\dagger, c_{j\downarrow}^\dagger, 
c_{j\uparrow}, c_{j\downarrow}$ are Fermi operators.
The Hubbard Hamilton operator commutes with the total number operator 
$\hat N$ and the total spin operator $\hat S_z$ where
$$
\hat N := \sum_{j=1}^2 (c_{j\uparrow}^\dagger c_{j\uparrow} + 
c_{j\downarrow}^\dagger c_{j\downarrow} ), \qquad
\hat S_z := \frac12 \sum_{j=1}^2 (c_{j\uparrow}^\dagger c_{j\uparrow}
- c_{j\downarrow}^\dagger c_{j\downarrow})\,.
$$
We consider the subspace with two particles $N=2$ and total spin
$S_z=0$. A basis in this four dimensional Hilbert space is given by
$$
c_{1\uparrow}^\dagger c_{1\downarrow}^\dagger|{\bf 0}\rangle, \quad
c_{1\uparrow}^\dagger c_{2\downarrow}^\dagger|{\bf 0}\rangle, \quad
c_{2\uparrow}^\dagger c_{1\downarrow}^\dagger|{\bf 0}\rangle, \quad
c_{2\uparrow}^\dagger c_{2\downarrow}^\dagger|{\bf 0}\rangle\,.
$$
We find the matrix representation of $\hat H$ for this basis. 
Using the Fermi anti-commutation relations and 
$c_{j\uparrow}|{\bf 0}\rangle=0$, $c_{j\downarrow}|{\bf 0}\rangle=0$
for $j=1,2$ we obtain the matrix representation of $\hat H$ with the given basis
$$
H = \pmatrix{ U & t & t & 0 \cr t & 0 & 0 & t \cr
              t & 0 & 0 & t \cr 0 & t & t & U }\,.
$$
We find the four permutation matrices $P_0=I_4=I_2 \star I_2=
I_2 \otimes I_2$,
$$
P_1 = I_2\star\pmatrix{ 0 & 1 \cr 1 & 0 },\quad
P_2 = \pmatrix{ 0 & 1 \cr 1 & 0 }\star I_2,\quad
P_3 = \pmatrix{ 0 & 1 \cr 1 & 0 }\star \pmatrix{ 0 & 1 \cr 1 & 0 }
$$
where we define the star product $\star$ of the two $2 \times 2$
matrices $A$, $B$ as \cite{6}
$$
\pmatrix{ a_{11} & a_{12} \cr a_{21} & a_{22} }
\star \pmatrix{ b_{11} & b_{12} \cr b_{21} & b_{22} }
:= \pmatrix{ a_{11} & 0 & 0 & a_{12} \cr 0 & b_{11} & b_{12} & 0 \cr
             0 & b_{21} & b_{22} & 0 \cr
             a_{21} & 0 & 0 & a_{22} }.
$$
We note that the star product of two $2 \times 2$ permutation
matrices is a $4 \times 4$ permutation matrix.
Here $P_1$ is the swap gate. The four permutation matrices 
$P_0$, $P_1$, $P_2$, $P_3$ form a commutative 
group with $P_j^2=I_4$ for $j=0,1,2,3$. If $P$ is an $n \times n$ 
permutation matrix with $P^2=I_n$ then
$$
\Pi_1 = \frac12 (I_n + P), \qquad \Pi_2 = \frac12 (I_n - P)
$$
are projection matrices with $\Pi_1\Pi_2=0_n$, where
$0_n$ is the $n \times n$ zero matrix. Using
the permutation matrices $P_0$ and $P_3$ (which form a subgroup) 
these projection operators can now be used to find the invariant 
subspaces
$$
\left\{ \, \pmatrix { 1 \cr 0 \cr 0 \cr 1 }, \,\,\,
\pmatrix { 0 \cr 1 \cr 1 \cr 0 } \, \right\}, \qquad
\left\{ \, \pmatrix { 1 \cr 0 \cr 0 \cr -1 }, \,\,\,
\pmatrix { 0 \cr 1 \cr -1 \cr 0 } \, \right\}\,.
$$
These four vectors (after normalization) are known in quantum 
computing as the Bell basis \cite{1,2,3,4,5}.
This leads to the two invariant sub Hilbert spaces
$$
\left\{\frac1{\sqrt2} 
(c_{1\downarrow}^\dagger c_{1\uparrow}^\dagger |0\rangle + 
c_{2\downarrow}^\dagger c_{2\uparrow}^\dagger |0\rangle), \quad
\frac{1}{\sqrt2} 
(c_{1\downarrow}^\dagger c_{2\uparrow}^\dagger |0\rangle +
c_{2\downarrow}^\dagger c_{1\uparrow}^\dagger |0\rangle) 
\right\}
$$
$$
\left\{\frac1{\sqrt2} 
(c_{1\downarrow}^\dagger c_{1\uparrow}^\dagger |0\rangle -
c_{2\downarrow}^\dagger c_{2\uparrow}^\dagger |0\rangle), \quad
\frac{1}{\sqrt2}
(c_{1\downarrow}^\dagger c_{2\uparrow}^\dagger |0\rangle -
c_{2\downarrow}^\dagger c_{1\uparrow}^\dagger |0\rangle) 
\right\}\,.
$$
Example 3. Let $\sigma_1$, $\sigma_2$, $\sigma_3$ be the Pauli spin
matrices
$$
\sigma_1 := \pmatrix {0 & 1 \cr 1 & 0}, \qquad
\sigma_2 := \pmatrix {0 & -i \cr i & 0}, \qquad
\sigma_3 := \pmatrix {1 & 0 \cr 0 & -1} .
$$
Consider the Hamilton operators \cite{7} 
$$
\hat H = \hbar\omega_1 \sigma_3 \otimes I_2 + \hbar\omega_2
I_2 \otimes \sigma_1 + \epsilon(\sigma_3 \otimes \sigma_1)
$$
$$
\hat K = \hbar\omega_1 \sigma_3 \otimes I_2 + \hbar\omega_2
I_2 \otimes \sigma_1 + \epsilon(\sigma_1 \otimes \sigma_3)
$$
where for the second Hamilton operator $\hat K$ the interaction
term is swapped around, i.e. $\sigma_3 \otimes \sigma_1 \to
\sigma_1 \otimes \sigma_3$. This provides symmetry breaking.
For the Hamilton operator $\hat H$ we find the symmetries (permutation
matrices)
$$
P_0 = I_4, \quad P_1 = I_2 \oplus \sigma_1, \quad
P_2 = \sigma_1 \oplus I_2, \quad P_3 = \sigma_1 \oplus \sigma_1 
$$
where $\oplus$ denotes the direct sum. The four matrices
form a commutative group under matrix multiplication.
All satisfy $P_j^2=I_4$. Thus we can use the projection
matrices $\Pi_1=(I_4+P_j)/2$, $\Pi_2=(I_4-P_j)/2$ to decompose
the Hilbert space into two invariant subspaces. 
On the other hand for the Hamilton operator $\hat K$ we only find 
the identity matrix $P_0=I_4$, i.e. no non-trivial symmetry is admitted.
\newline

Example 4. Consider the Hamilton operator for triple spin
interaction
$$ 
\hat H = \sigma_1 \otimes \sigma_2 \otimes \sigma_3\,. 
$$
The eigenvalues of this hermitian and unitary $8 \times 8$
matrix are $+1$ (four-fold degenerate) and $-1$ (four-fold degenerate).
Owing to this degeneracy one expects a ``large'' number of symmetries.
Applying the SymbolicC++ code we find 24 permutation matrices
listed $P_0, P_1,\dots,P_{23}$ with $P_0=I_8$. They form
a non-commutative group under matrix multiplication and are a subgroup
of the group of $8 \times 8$ permutation matrices.
We note that the Kronecker product $\otimes$ and the direct
sum $\oplus$ of two permutation matrices is again a 
permutation matrix \cite{8}. 
Now we can list the ones with $P_j^2=I_8$. We have
\begin{eqnarray*}
P_1 &=& I_2 \oplus \left(\pmatrix { 0 & 1 \cr 1 & 0 }
\otimes \pmatrix { 0 & 1 \cr 1 & 0 }\right) \oplus I_2 \\
P_2 &=& (I_2\star\sigma_1)\oplus(\sigma_1\star I_2) \\
P_5 &=& I_2\otimes I_2\otimes\pmatrix{1&0\cr 0&0}
      + \sigma_1\otimes I_2\otimes\pmatrix{0&0\cr 0&1} \\
P_6 &=& (\sigma_1\star I_2)\oplus(I_2\star\sigma_1) \\
P_8 &=& I_2\otimes\sigma_1\otimes\sigma_1 \\
P_{13} &=& I_2\otimes I_2\otimes\pmatrix{0&0\cr 0&1}
       + \sigma_1\otimes I_2\otimes\pmatrix{1&0\cr 0&0} \\
P_{15} &=& \sigma_1\otimes I_2\otimes I_2 \\
P_{23} &=& \sigma_1 \otimes \sigma_1 \otimes \sigma_1\,.
\end{eqnarray*}
The other ones can be found by multiplication of these 
permutation matrices, for example $P_3=P_1P_2$ etc.
Thus the 24 matrices form a subgroup of the permutation 
group of $8 \times 8$ matrices.
\newline

Another spin Hamilton operator studied is \cite{9}
$$
\hat H = a\sum_{j=1}^4 \sigma_3(j) \sigma_3(j+1) + b\sum_{j=1}^4 \sigma_1(j)
$$
with cyclic boundary conditions, i.e. $\sigma_3(5) \equiv \sigma_3(1)$.
Here $a, b$ are real constants and $\sigma_1$, $\sigma_2$
and $\sigma_3$ are the Pauli matrices. Thus the underlying Hilbert space
is ${\mathbb C}^{16}$. Recall that
$$
\sigma_k(1) = \sigma_k \otimes I_2 \otimes I_2 \otimes I_2, \qquad
\sigma_k(2) = I_2 \otimes \sigma_k \otimes I_2 \otimes I_2
$$
$$
\sigma_k(3) = I_2 \otimes I_2 \otimes \sigma_k \otimes I_2, \qquad
\sigma_k(4) = I_2 \otimes I_2 \otimes I_2 \otimes \sigma_k
$$
where $k=1,2,3$. We obtain the symmetric $16 \times 16$ matrix for $\hat H$
$$
\pmatrix
{4a & b & b & 0 & b & 0 & 0 & 0 & b & 0 & 0 & 0 & 0 & 0 & 0 & 0\cr
  b & 0 & 0 & b & 0 & b & 0 & 0 & 0 & b & 0 & 0 & 0 & 0 & 0 & 0\cr
  b & 0 & 0 & b & 0 & 0 & b & 0 & 0 & 0 & b & 0 & 0 & 0 & 0 & 0\cr
  0 & b & b & 0 & 0 & 0 & 0 & b & 0 & 0 & 0 & b & 0 & 0 & 0 & 0\cr
  b & 0 & 0 & 0 & 0 & b & b & 0 & 0 & 0 & 0 & 0 & b & 0 & 0 & 0\cr
  0 & b & 0 & 0 & b & -4a & 0 & b & 0 & 0 & 0 & 0 & 0 & b & 0 & 0\cr
  0 & 0 & b & 0 & b & 0 & 0 & b & 0 & 0 & 0 & 0 & 0 & 0 & b & 0\cr
  0 & 0 & 0 & b & 0 & b & b & 0 & 0 & 0 & 0 & 0 & 0 & 0 & 0 & b\cr
  b & 0 & 0 & 0 & 0 & 0 & 0 & 0 & 0 & b & b & 0 & b & 0 & 0 & 0\cr
  0 & b & 0 & 0 & 0 & 0 & 0 & 0 & b & 0 & 0 & b & 0 & b & 0 & 0\cr
  0 & 0 & b & 0 & 0 & 0 & 0 & 0 & b & 0 & -4a & b & 0 & 0 & b & 0\cr
  0 & 0 & 0 & b & 0 & 0 & 0 & 0 & 0 & b & b & 0 & 0 & 0 & 0 & b\cr
  0 & 0 & 0 & 0 & b & 0 & 0 & 0 & b & 0 & 0 & 0 & 0 & b & b & 0\cr
  0 & 0 & 0 & 0 & 0 & b & 0 & 0 & 0 & b & 0 & 0 & b & 0 & 0 & b\cr
  0 & 0 & 0 & 0 & 0 & 0 & b & 0 & 0 & 0 & b & 0 & b & 0 & 0 & b\cr
  0 & 0 & 0 & 0 & 0 & 0 & 0 & b & 0 & 0 & 0 & b & 0 & b & b & 4a}\,.
$$
The Hamilton operator $\hat H$ admits the $C_{4v}$
symmetry group. The order of this non-commutative group
is 8. One finds the following set of eight symmetries \cite{9}
$$
E : (1,2,3,4) \to (1,2,3,4) \qquad C_2 : (1,2,3,4) \to (3,4,1,2)
$$
$$
C_4 : (1,2,3,4) \to (2,3,4,1) \qquad C_4^3 : (1,2,3,4) \to (4,1,2,3)
$$
$$
\sigma_v : (1,2,3,4) \to (2,1,4,3) \qquad \sigma_v' : (1,2,3,4) \to (4,3,2,1)
$$
$$
\sigma_d : (1,2,3,4) \to (1,4,3,2) \qquad \sigma_d' : (1,2,3,4) \to (3,2,1,4)
$$
which form a group isomorphic to $C_{4v}$. 
The symmetries can be found by calculating the $16 \times 16$ 
permutation matrices such that $\hat H=P^T\hat HP$.

\section{Code Description}

Algorithms for finding all permutations of a sequence of objects
are described by Knuth \cite{10}. For a given $n$ the permutation matrices
are generated with the following algorithm. The algorithm implements
the nested loops\\

\hspace*{1ex} For $j_0=0,1,\ldots,n-1$ do\\
\hspace*{2ex}   For $j_1=0,1,\ldots,n-1$ do\\
\hspace*{4ex}       \rotatebox{-30}{$\ddots$}\\
\hspace*{4ex}     For $j_{n-1}=0,1,\ldots,n-1$ do\\
\hspace*{5ex}         If $j_0\neq j_1\neq\cdots\neq j_{n-1}$ then\\
\hspace*{6ex}           use the permutation
                        $(0,1,2,\ldots,n-1)\to(j_0,j_1,j_2,\ldots,j_{n-1})$.\\
\hspace*{4ex}     End loop\\
\hspace*{4ex}       \rotatebox{110}{$\ddots$}\\
\hspace*{3ex}   End loop\\
\hspace*{1ex} End loop\\

\paragraph*{Algorithm to find all permutation matrices.}

\begin{enumerate}
 \item Create an array $(j_0,j_1,\ldots,j_{n-1})$
       of loop variables.
 \item Initialize $j_k:=-1$ for $k=0,1,\ldots,n-1$.
 \item Initialize the loop variable index $i$ to $i:=0$.
 \item \label{loop0}%
       While $i\geq 0$
       \begin{enumerate}
        \item \textit{Iterate.}\\
              Set $j_i:= j_i+1$.
        \item \label{loop1}%
              \textit{Termination condition.}\\
              If $j_i = n$ terminate this loop:
              \begin{enumerate}
               \item Set $j_i:=-1$.
               \item \textit{Exit the nested loop.}\\
                     Set $i:= i-1$.
               \item Goto \ref{loop0}.
              \end{enumerate}
        \item If $j_k=j_i$ for some $k\in\{0,1,\ldots,i-1\}$
              then goto \ref{loop0}.
        \item \textit{Enter the next nested loop.}\\
              Set $i:=i+1$.
        \item \textit{Innermost loop completes.}\\
              If $i=n$ then use the permutation
              $(0,1,2,\ldots,n-1)\to(j_0,j_1,j_2,\ldots,j_{n-1})$
              i.e. the permutation matrix $P$ is given by
              $$(P)_{uv}=\left\{\begin{array}{ll}
                                 1& \textnormal{if $v=j_u$}\\
                                 0& \textnormal{otherwise}
                                \end{array}\right..$$
       \end{enumerate}
\end{enumerate}
The SymbolicC++ program \cite{11} utilizes the vector class of the 
Standard Template Library. The Hamilton operator refers to example
2 in the text (two point Hubbard model).

\small
\begin{verbatim}
// permutation.cpp

#include <iostream>
#include <vector>
#include "symbolicc++.h"
using namespace std;

int total;
Symbolic H;

void commutes(const Symbolic &P)
{
 if(P*H==H*P) cout << "P[" << total++ << "] = " << P << endl;
}

void find_perm(int n,void (*use)(const Symbolic&))
{
 int i, k;
 Symbolic P;
 vector<int> j(n,-1);

 i = 0; j[0] = -1;
 while(i >= 0)
 {
  if(++j[i]==n) { j[i--] = -1; continue; }
  if(i < 0) break;
  for(k=0;k<i;++k) if(j[k]==j[i]) break;
  if(k!=i) continue;
  ++i;
  if(i==n)
  {
  P = 0*Symbolic("",n,n);
  for(k=0;k<n;k++) P(k,j[k]) = 1;
  use(P);
  --i;
  }
 }
}

int main(void)
{
 using SymbolicConstant::i;
 Symbolic sqrt2 = sqrt(Symbolic(2));
 Symbolic U("U"); Symbolic t("t");
 H = ((U,t,t,Symbolic(0)),(t,Symbolic(0),Symbolic(0),t),
      (t,Symbolic(0),Symbolic(0),t),(Symbolic(0),t,t,U));
 cout << "H = " << H << endl;
 total = 0;
 find_perm(H.rows(),commutes);
 return 0;
}
\end{verbatim}
\normalsize

A Maxima implementation is available from the authors.

\section{Conclusion}

We applied a brute force method to find all possible permutation
matrices that provide symmetries for given Hamilton
operators in a finite dimensional Hilbert space.
With growing size of the Hamilton operators
matrix representation finding
the permutation matrices becomes very time-consuming.
A more efficient approach would be to find only the generators
of the group of permutation matrices that provide symmetries for
a given Hamilton operator. Another open question is how this method 
can be extended to find other classes of symmetries.
\newline

{\bf Acknowledgment}
\newline

The authors are supported by the National Research Foundation (NRF),
South Africa. This work is based upon research supported by the National
Research Foundation. Any opinion, findings and conclusions or recommendations
expressed in this material are those of the author(s) and therefore the
NRF do not accept any liability in regard thereto.

\strut\hfill

\end{document}